\documentclass[jkps,preprint,showpacs,floatfix]{revtex4}
\usepackage[dvips]{graphicx}
\begin{document}
\title{$N_pN_n$ scheme and the valence proton-neutron interaction}
\author{Youngshin \surname{Byun}}
\author{Guanghao \surname{Jin}}
\author{Jin-Hee \surname{Yoon}}
\author{Dongwoo \surname{Cha}}
\email{dcha@inha.ac.kr}
\thanks{Fax: +82-32-866-2452}
\affiliation{Department of Physics, Inha University, Incheon
402-751, Korea}
\date{May 19, 2008}

\begin{abstract}
We examine the common belief that the $N_pN_n$ scheme is manifested as a direct consequence of the valence proton-neutron interaction which has proven to be a dominant factor in developing collectivity in nuclei. We show that the simplification of the $N_pN_n$-plot of the lowest $2^+$ excitation energy is introduced merely because the excitation energy always decreases when the valence nucleon number becomes larger.
\end{abstract}


\maketitle

Since de-Shalit and Goldhaber recognized a critical role of the proton-neutron (p-n) interaction in developing mixed configurations in nuclei half a century ago \cite{deShalit}, many authors have asserted the importance of the valence p-n interaction in the evolution of nuclear structure. Talmi was the first to emphasize that the p-n interaction may give rise to deformed nuclei \cite{Talmi}.  Subsequently, Federman and Pittel have shown explicitly, by microscopic calculations, that nuclear deformation is produced by the isoscalar component of the p-n interaction between nucleons in spin-orbit partner orbits \cite{Federman}.

\begin{figure}[b]
\centering
\includegraphics[width=14.0cm,angle=0]{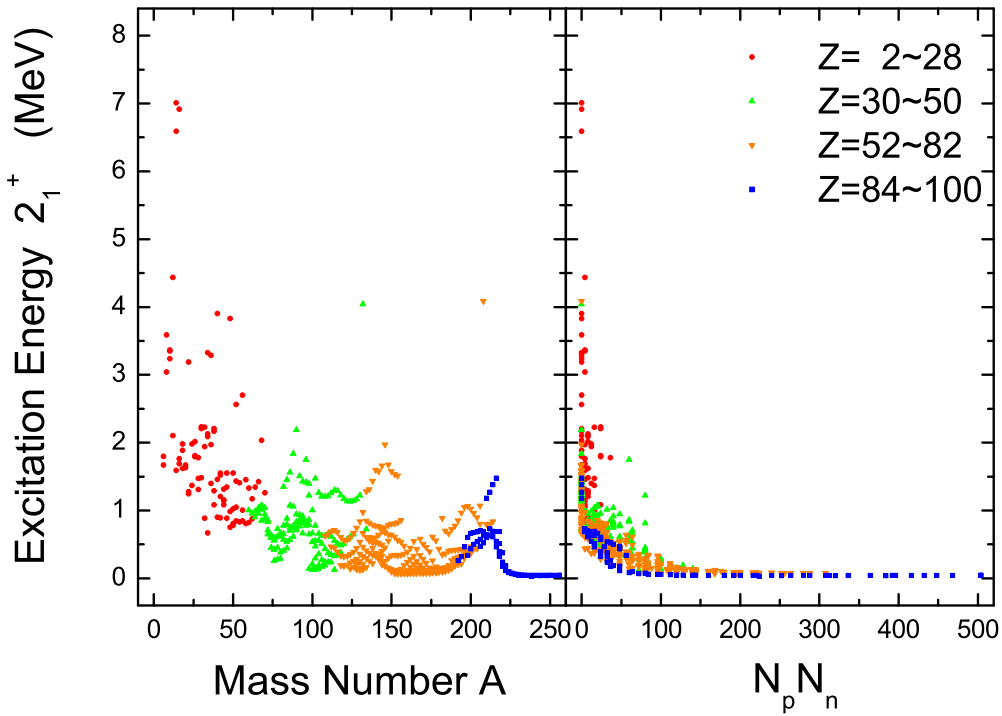}
\caption{A typical example demonstrating the $N_pN_n$ scheme. Measured excitation energies of the lowest $2^+$ states in even-even nuclei are plotted against the mass number $A$ in the left panel and against the product $N_pN_n$ in the right panel. The excitation energies are quoted from Ref.\,\cite{Raman}.}
\label{fig-1}
\end{figure}

Meanwhile, Casten noticed that a simple pattern appeared whenever nuclear data concerning nuclear deformation was plotted against the product $N_pN_n$ between the valence proton number $N_p$ and the valence neutron number $N_n$ \cite{Casten1}. This phenomenon has been referred to as ``the $N_pN_n$ scheme'' in the literature \cite{Casten2}. For a typical example of the $N_pN_n$ scheme, consider the graphs shown in the two panels of Fig.\,\ref{fig-1} \cite{Yoon}. When the measured excitation energies $E_x(2_1^+)$ of the lowest $2^+$ states in even-even nuclei are plotted against the mass number $A$ ($A$-plot), we get data points scattered irregularly over the $E_x-A$ plane as seen in the left panel of Fig.\,\ref{fig-1}. But when the same data points are plotted against the product $N_pN_n$ ($N_pN_n$-plot), as shown in the right panel of Fig.\,\ref{fig-1}, we suddenly find that the data points are neatly rearranged. A similar simplification was also observed from the data on the observables which were related to nuclear deformation such as the excitation energy ratio $E_x(4_1^+)/E_x(2_1^+)$ \cite{Casten3,Casten4,Cakirli}, the transition probability $B(E2;2_1^+ \rightarrow 0^+)$ \cite{Casten5}, and the quadrupole deformation parameter $e_2$ \cite{Zhao}.

Casten took the lead, more than two decades ago, in regarding the $N_pN_n$ scheme as clear evidence of the p-n interaction being the dominant factor of inducing the nuclear deformation \cite{Casten3}. Since then, in almost all of the published work related to that subject, it has been taken for granted that the $N_pN_n$ scheme is manifested as a direct consequence of the valence p-n interaction \cite{Espino,Gupta,Saha,Zhao2}. However, it seems to us that there is no direct proof of the $N_pN_n$ scheme appearing due to the underling p-n interaction among the valence nucleons. All we have is the following two confirmed statements: (i) The nuclear deformation is induced by the p-n interaction, and (ii) the $N_pN_n$ scheme is observed for the data dealing with nuclear deformation. Even though it is true that there is a close relation between the nuclear deformation and the p-n interaction and also between the $N_pN_n$ scheme and the nuclear deformation, the above two statements do not automatically guarantee that the $N_pN_n$ scheme comes about due to the valence p-n interaction. In this work, therefore, we want to examine whether there really exists a causal relationship between the $N_pN_n$ scheme and the p-n interaction.

\begin{figure}[t]
\centering
\includegraphics[width=14.0cm,angle=0]{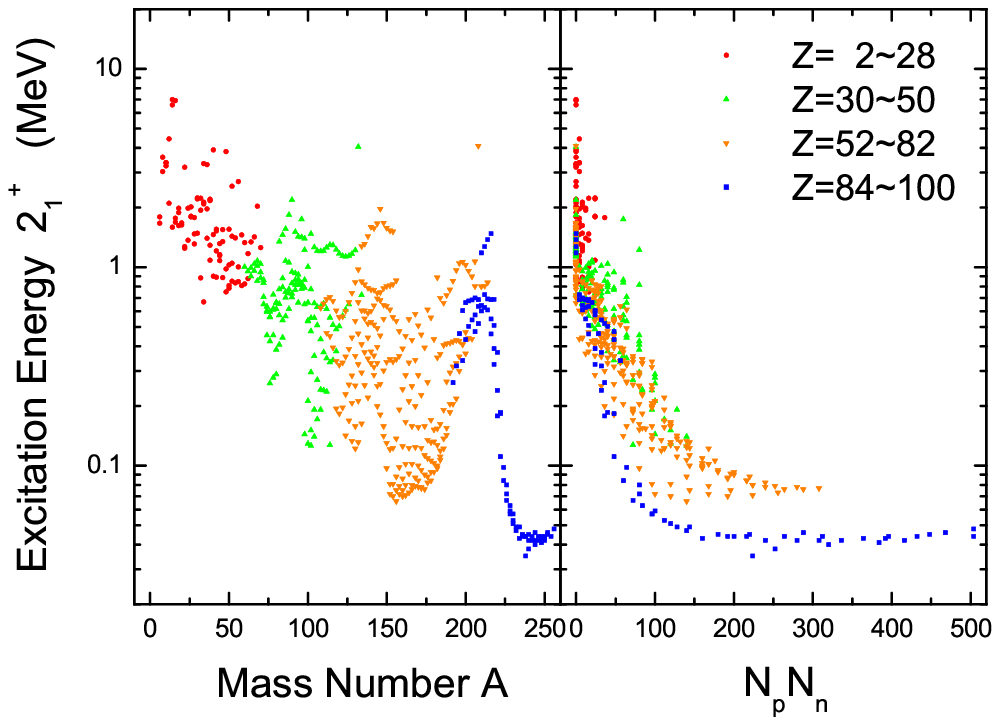}
\caption{Same as in Fig.\,\ref{fig-1} but for the logarithmic vertical scale instead of the linear one.}
\label{fig-2}
\end{figure}

\begin{figure}[b]
\centering
\includegraphics[width=14.0cm,angle=0]{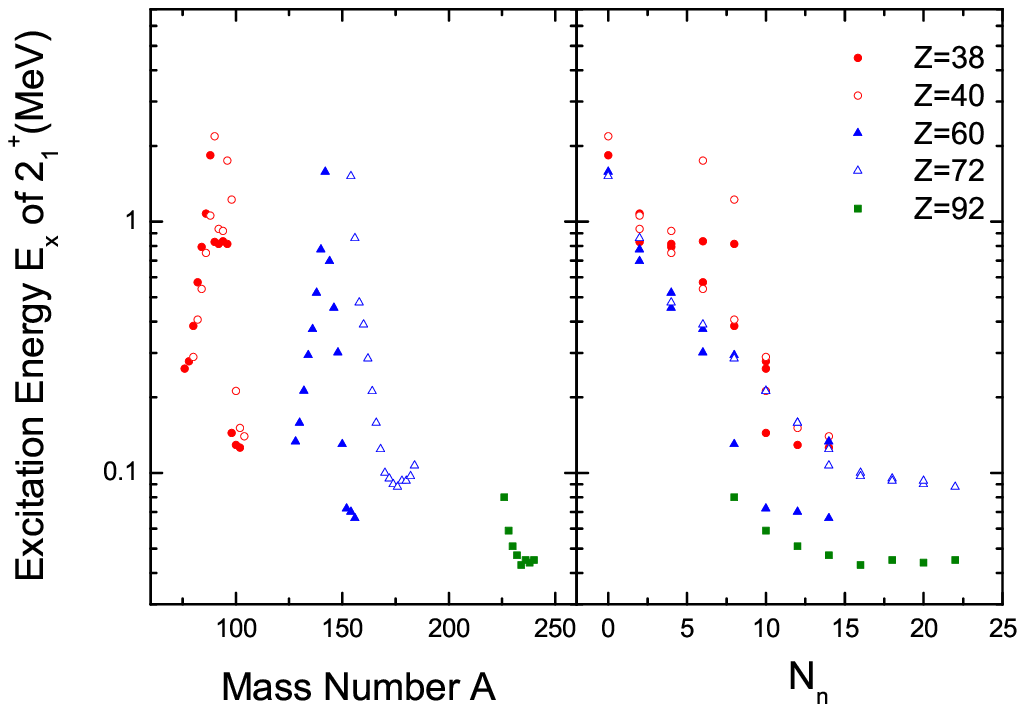}
\caption{Measured excitation energies of the lowest $2^+$ state in isotopes whose valence proton number $N_p$ is equal to $10$. In the left panel, the excitation energies are plotted against the mass number $A$, while in the right panel, they are plotted against the valence neutron number $N_n$.}
\label{fig-3}
\end{figure}

Let us start with Fig.\,\ref{fig-1} where we depict the lowest $2^+$ excitation energies by the following different symbols according to which proton major shell they belong to: solid circles ($Z=2 \sim 28$); upward solid triangles ($Z=30 \sim 50$); downward solid triangles ($Z=52 \sim 82$); and solid squares ($Z=84 \sim 100$). From this figure, we can easily observe that while the data points in the $A$-plot (left panel) occupy a particular range of the mass number $A$ exclusively depending on which proton major shell they belong to, the same data points in the $N_pN_n$-plot (right panel) always occupy from zero up to a certain maximum number of the product $N_pN_n$. The way that the data points in the $A$-plot are relocated in the $N_pN_n$-plot can be more clearly seen from Fig.\,\ref{fig-2}. It is exactly the same graph as the one shown in Fig.\,\ref{fig-1} except employing a logarithmic vertical scale instead of the linear one. We find from the $N_pN_n$-plot (right panel) of Fig.\,\ref{fig-2} that the excitation energies $E_x(2_1^+)$ belonging to  each proton major shell rearrange themselves in the $E_x-N_pN_n$ plane separately in such a way that they decrease monotonically as $N_pN_n$ becomes larger. This means that the data points from each and every proton major shell actually reveal individually the characteristic feature of the $N_pN_n$ scheme, namely the fact that a simple pattern appears whenever the $N_pN_n$-plot is drawn.

However, we want to provide evidence showing that the monotonically decreasing simple pattern of the excitation energies $E_x(2_1^+)$ found in the $N_pN_n$-plot originates not from some kind of nuclear p-n correlations but simply from the characteristic feature of $E_x(2_1^+)$ which depends only on $N_p$ and $N_n$ separately. In Fig.\,\ref{fig-3}, we show the measured excitation energies of the lowest $2^+$ states in isotopes whose atomic numbers are $Z=38$(solid circles), $Z=40$(open circles), $Z=60$(solid triangles), $Z=72$(open triangles), and $Z=92$(solid squares). Note that these isotopes all have the same valence proton number, $N_p=10$. In the left and right panels of this figure, $E_x(2_1^+)$ is plotted against the mass number $A$ ($A$-plot) and the valence neutron number $N_n$ ($N_n$-plot), respectively. Now we can easily find that $E_x(2_1^+)$, belonging to the same isotope, sometimes increases and sometimes decreases in the $A$-plot, but it always decreases in the $N_n$-plot. Actually, this property of $E_x(2_1^+)$ decreasing monotonically with $N_n$ for fixed $N_p$, prevails in all isotopes without exception for the entire chart of nuclides. In addition, exactly the same sort of graph is obtained when $E_x(2_1^+)$, belonging to the same isotones, is plotted against $N_p$ for fixed $N_n$. This is precisely the reason why $E_x(2_1^+)$ decreases monotonically with $N_pN_n$ in the $N_pN_n$-plot.

\begin{figure}[b]
\centering
\includegraphics[width=14.0cm,angle=0]{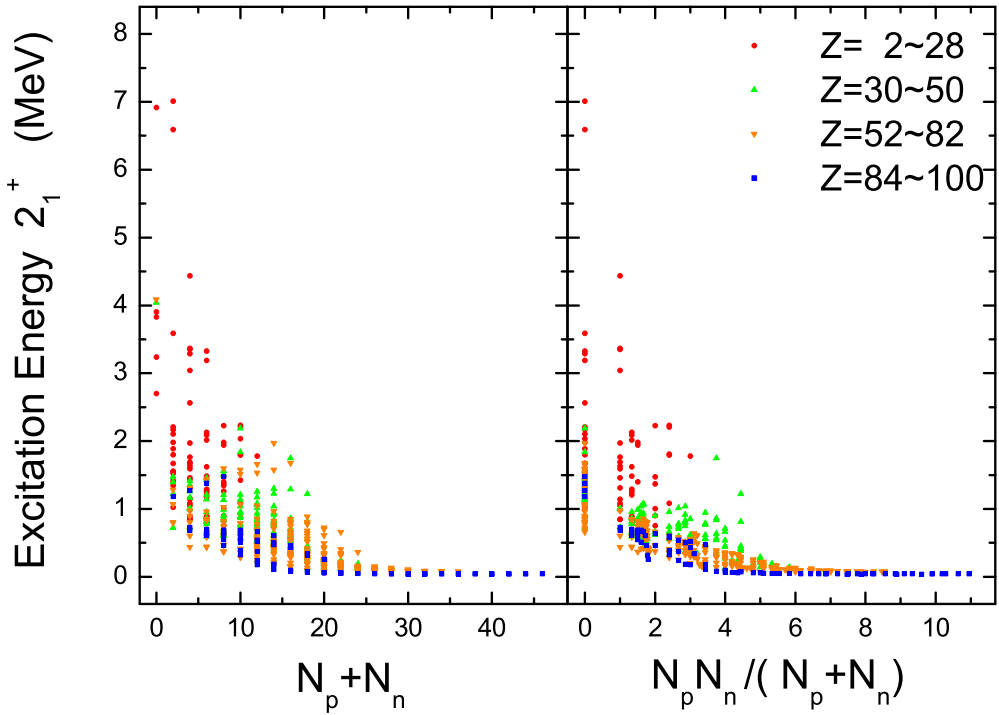}
\caption{Measured excitation energies of the lowest $2^+$ states in even-even nuclei are plotted against the sum $N_p+N_n$ in the left panel and against the Casten factor $P=N_pN_n/(N_p+N_n)$ in the right panel.}
\label{fig-4}
\end{figure}

Therefore, we say that attributes claimed by the $N_pN_n$ scheme are not, in fact, inherent only to the $N_pN_n$-plot. For example, any plot, like the one shown in Fig.\,\ref{fig-4}, against a quantity, which increases with $N_p$ or $N_n$, shows a similar tendency as the $N_pN_n$-plot. The excitation energies $E_x(2_1^+)$ are depicted against the sum $N_p+N_n$ (($N_p+N_n$)-plot) in the left panel of Fig.\,\ref{fig-4} while they are plotted again against the Casten factor $P=N_pN_n/(N_p+N_n)$ ($P$-plot) in the right panel of the same figure. By comparing the two plots (the ($N_p+N_n$)-plot and the $P$-plot) of Fig.\,\ref{fig-4} with the $N_pN_n$-plot, shown in the right panel of Fig.\,\ref{fig-1}, we see that the overall shape is quite similar except for the lower left corner of the graph. It is quite evident that the breadth of this part of the $N_pN_n$-plot shown in Fig.\,\ref{fig-1} is considerably narrower than that of the ($N_p+N_n$)-plot or the $P$-plot shown in Fig.\,\ref{fig-4}.

\begin{figure}[t]
\centering
\includegraphics[width=14.0cm,angle=0]{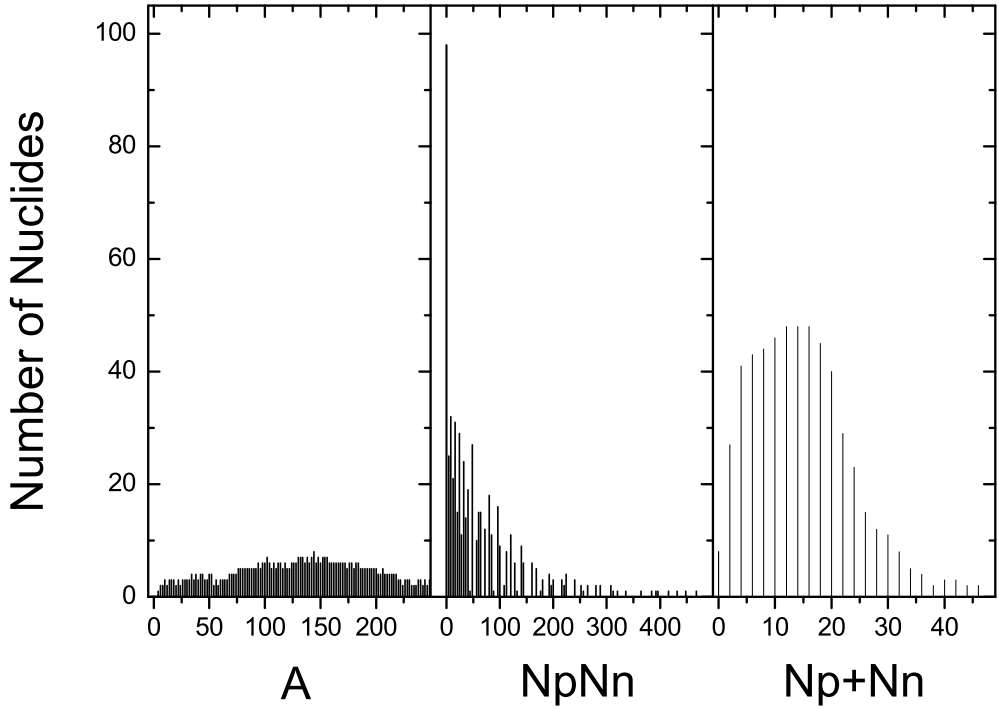}
\caption{Number of nuclides where $2^+$ excited states are observed in even-even nuclei against the mass number $A$, the product $N_pN_n$, and the sum $N_p+N_n$.}
\label{fig-5}
\end{figure}

However, once again, we argue that the difference in breadth of the mentioned various plots does not arise from some kinds of dynamical reasons but simply from the difference in multiplicity profile of nuclides as can be seen from Fig.\,\ref{fig-5}. In this figure, the number of nuclides where $2^+$ excited states are observed in even-even nuclei is plotted against the mass number $A$ (left panel, the $A$-multiplicity), the product $N_pN_n$ (central panel, the $N_pN_n$-multiplicity), and the sum $N_p+N_n$ (right panel, the ($N_p+N_n$)-multiplicity). According to this figure, the $A$-multiplicity is more or less constant for all $A$ and has a small value of less than 10. In contrast to the $A$-multiplicity, the $N_pN_n$ multiplicity decreases steeply from the maximum value of around 100 at $N_pN_n$=0, while the ($N_p+N_n$)-multiplicity stays around 50 up to $N_p+N_p=20$. We contend that this very difference between the $N_pN_n$-multiplicity and the ($N_p+N_n$)-multiplicity must be the most direct origin of the difference of breadth between the $N_pN_n$-plot and the ($N_p+N_n$)-plot. More specifically, the data points, up to $N_pN_n$=100 in the $N_pN_n$-plot, are stretched to make those up to $N_p+N_n=20$ in the ($N_p+N_n$)-plot. Therefore, it may not be reasonable to insist that the p-n interaction plays an important role just because the $N_pN_n$-plot shows a simple pattern or that the p-n interaction is not important just because the ($N_p+N_n$)-plot shows a simple pattern.

\begin{figure}[t]
\centering
\includegraphics[width=14.0cm,angle=0]{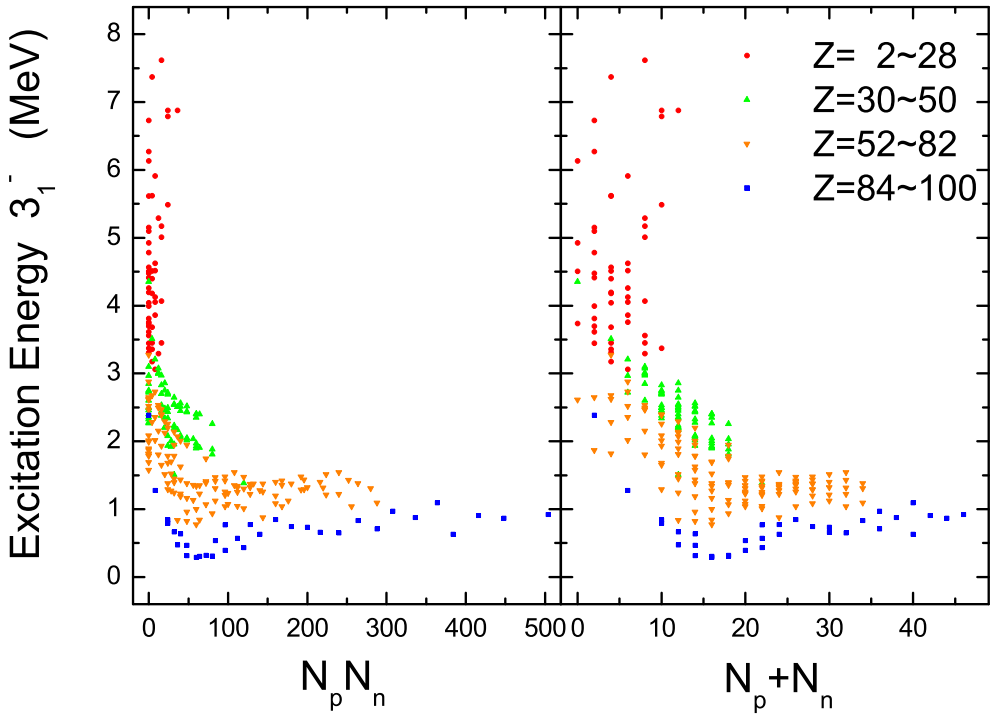}
\caption{Measured excitation energies of the lowest $3^-$ states in even-even nuclei are plotted against the product $N_pN_n$ in the left panel and against the sum $N_p+N_n$ in the right panel. The excitation energies are quoted from Ref.\,\cite{Kibedi}.}
\label{fig-6}
\end{figure}

For example, Zamfir {\it et al}. showed that the excitation energy $E_x(3_1^-)$ of the lowest $3^-$ state in even-even nuclei could be well parametrized by the sum $N_p+N_n$ \cite{Zamfir}. Based on the argument that the pairing interaction is the dominant mechanism for the low-lying octupole excitations, they proceeded to regard the $N_p+N_n$ parametrization as evidence of separate contributions to $E_x(3_1^-)$ from each kind of nucleon. In order to elucidate this problem, we plotted the measured excitation energies $E_x(3_1^-)$ against the product $N_pN_n$ (left panel) as well as against the sum $N_p+N_n$ (right panel) in Fig.\,\ref{fig-6}. As is evident from Fig.\,\ref{fig-6}, the ($N_p+N_n$)-plot of $E_x(3_1^-)$ is nothing but the stretched version of the $N_pN_n$-plot of $E_x(3_1^-)$ just like the $E_x(2_1^+)$ case explained in the previous paragraph regarding the ($N_p+N_n$)-plot of $E_x(2_1^+)$ in Fig.\,\ref{fig-4}. Consequently, if we can find a good parametrization of $N_p+N_n$ from the lowest octupole excitation energy data, we must also be able to find an equivalently good parametrization of $N_pN_n$ from the same data. Therefore, it is not wise to conclude about the nature of the valence nucleon correlations merely based on the $N_pN_n$-plot or ($N_p+N_n$)-plot alone.

Incidentally, as for the valence nucleon number dependence of the lowest excitation energy, we mention a recently proposed simple empirical formula we developed where we demonstrated being able to describe the essential trends of $E_x(2_1^+)$ in even-even nuclei throughout the entire periodic table \cite{Ha,Jin1}. Subsequently, it was demonstrated that the same empirical formula can be employed to describe  $E_x(J_1^\pi)$ of the lowest natural parity states of all the multipoles $J^\pi$ up to $J=10^+$. The formula is given by \cite{Jin2}
\begin{equation} \label{E}
E_x (J_1^\pi) = \alpha(\pi) J^{a(\pi)} A^{-\gamma(\pi) J^{c(\pi)}}  + \beta_p(\pi)  e^{- {\lambda_p(\pi) N_p \over \sqrt{J}}}  + \beta_n(\pi) e^{- {\lambda_n(\pi) N_n \over \sqrt{J}}}
\end{equation}
where the parity $\pi$ dependent model parameters $\alpha(\pi)$, $a(\pi)$, $\gamma(\pi)$, $c(\pi)$, $\beta_p(\pi)$, $\lambda_p(\pi)$, $\beta_n(\pi)$, and $\lambda_n(\pi)$ are fitted from the data. (For specific parameter values, see Ref.\,\cite{Jin2}.) Furthermore, we also showed that this formula not only complies with the $N_pN_n$ scheme but also has a composition which is ideal for revealing the $N_pN_n$ scheme \cite{Yoon}. Therefore, this empirical formula can be taken as another piece of evidence that the $N_pN_n$ scheme alone does not guarantee the importance of the valence p-n interaction.

In short, we have shown explicitly that the simple pattern observed from the $N_pN_n$-plot of $E_x(2_1^+)$ actually originates from the property that it decreases monotonically with $N_n$ for fixed $N_p$ and also with $N_p$ for fixed $N_n$. The $N_pN_n$ scheme manifested by other observables can also be explained similarly. For example, the $B(E2)$ values increase monotonically with $N_n$ for fixed $N_p$ and also with $N_p$ for fixed $N_n$. This means, in turn, that attributes claimed by the $N_pN_n$ scheme are not, in fact, inherent only to the $N_pN_n$-plot. Therefore, it is not wise to conclude that the p-n interaction plays an important role just because the $N_pN_n$-plot shows a simple pattern. Also we mention the empirical formula for $E_x(J_1^\pi)$, given by Eq.\,(\ref{E}), which may provide an appropriate cause why the $N_pN_n$-plot of $E_x(2_1^+)$ becomes so simple.

\begin{acknowledgments}
This work was supported by an Inha University research grant.
\end{acknowledgments}


\begin{references}
\bibitem{deShalit} A. de-Shalit and M. Goldhaber, Phys. Rev. 92 (1953) 1211.
\bibitem{Talmi} I. Talmi, Rev. Mod. Phys. 34 (1962) 704.
\bibitem{Federman} P. Federman and S. Pittel, Phys. Rev. C 20 (1979) 820.
\bibitem{Casten1} R.F. Casten, Nucl. Phys. A 443 (1985) 1.
\bibitem{Casten2} For a review of the $N_pN_n$ scheme, see R.F. Casten and N.V. Zamfir, J. Phys. G: Nucl. Part. Phys. 22 (1996) 1521.
\bibitem{Yoon} J.-H. Yoon, E. Ha, and D. Cha, J. Phys. G: Nucl. Part. Phys. 34 (2007) 2545.
\bibitem{Casten3} R.F. Casten, Phys. Rev. Lett. 54 (1985) 1991.
\bibitem{Casten4} R.F. Casten, D.S. Brenner, and P.E. Haustein, Phys. Rev. Lett. 58 (1987) 658.
\bibitem{Cakirli} R.B. Cakirli and R.F. Casten, Phys. Rev. Lett. 96 (2006) 132501.
\bibitem{Casten5} R.F. Casten and N.V.Zamfir, Phys. Rev. Lett. 70 (1993) 402.
\bibitem{Zhao} Y.M. Zhao, R.F. Casten, and A. Arima, Phys. Rev. Lett. 85 (2000) 720.
\bibitem{Raman} S. Raman, C.W. Nestor Jr., and P. Tikkanen, At.
    Data Nucl. Data Tables 78 (2001) 1.
\bibitem{Espino} J.M. Espino and J.D. Garrett, Nucl. Phys. A 492 (1989) 205.
\bibitem{Gupta} J.B. Gupta, H.M. Mittal, and J.H. Hamilton, A.V. Ramayya, Phys. Rev. C 42 (1990) 1373.
\bibitem{Saha} M. Saha and S. Sen, Phys. Rev. C 46 (1992) R1587.
\bibitem{Zhao2} Y.M. Zhao and A. Arima, Phys. Rev. C 68 (2003) 017301.
\bibitem{Kibedi} T. Kib{\' e}di and R.H. Spear, At. Data Nucl. Data Tables 80 (2002) 35.
\bibitem{Zamfir} N.V. Zamfir, R.F. Casten and P. Von Brentano, Phys. Lett. B 226 (1989) 11.
\bibitem{Ha} E. Ha and D. Cha, J. Korean Phys. Soc. 50 (2007) 1172.
\bibitem{Jin1} G. Jin, D. Cha and J.-H. Yoon, J. Korean Phys. Soc. 52 (2008) 1164.
\bibitem{Jin2} G. Jin, D. Cha and J.-H. Yoon, Preprint arXiv:0804.2527 [nucl-th] (2008).


\end{references}
\end{document}